\documentclass[
reprint,
superscriptaddress,
amsmath,amssymb,
showkeys,
aps,
prb,
]{revtex4-2}

\usepackage{threeparttable}
\usepackage{setspace}
\usepackage{makecell}
\usepackage{amsmath}
\usepackage{graphicx}
\usepackage{dcolumn}
\usepackage{bm}
\usepackage{hyperref}
\usepackage[mathlines]{lineno}
\usepackage{textcomp}
\usepackage{gensymb}
\usepackage{subfigure}
\usepackage{url}
\usepackage{xcolor}
\usepackage{hhline}
\usepackage[version=3]{mhchem}
\usepackage{multirow}
\usepackage[american]{babel}
\usepackage{hhline}
\usepackage{textgreek}
\usepackage{float} 
\usepackage{fixltx2e}

\usepackage{afterpage}

\definecolor{red}{rgb}{0.6,0.1,0.1}
\definecolor{blue}{rgb}{0.1,0.1,0.6}
\definecolor{green}{rgb}{0.1,0.6,0.1}
\newcommand{\FD}[1]{\textcolor{black}{#1}}
\newcommand{\HR}[1]{\textcolor{black}{#1}}

\newcommand{\bGaO}{$\beta$-$\ce{Ga2O3}$}
\newcommand{\gGaO}{$\gamma$-$\ce{Ga2O3}$}
\newcommand{\GaO}{$\ce{Ga2O3}$}
\newcommand{\IV}[2]{#1$_\text{#2}$}

\begin{document}

\title{Ultrahigh Stability of O-Sublattice in \texorpdfstring{\textit{\textbeta}-\ce{Ga2O3}}{}}

\author{Ru He}
\affiliation{Department of Physics, University of Helsinki, P.O. Box 43, FI-00014, Finland}

\author{Junlei Zhao} 
\affiliation{Department of Electrical and Electronic Engineering, Southern University of Science and Technology, Shenzhen 518055, China.}

\author{Jesper Byggm{\"a}star}
\affiliation{Department of Physics, University of Helsinki, P.O. Box 43, FI-00014, Finland}

\author{Huan He} 
\affiliation{School of Nuclear Science and Technology, Xi'an Jiaotong University, Xi'an, Shaanxi 710049, China.}
\affiliation{Department of Physics, University of Helsinki, P.O. Box 43, FI-00014, Finland}

\author{Flyura Djurabekova*} 
\affiliation{Department of Physics, University of Helsinki, P.O. Box 43, FI-00014, Finland}
\affiliation{Helsinki Institute of Physics, University of Helsinki, P.O. Box 43, FI-00014, Finland}


\begin{abstract}

Recently reported remarkably high radiation tolerance of $\gamma$/$\beta$-$\ce{Ga2O3}$ double-polymorphic structure 
brings this ultrawide bandgap semiconductor to the frontiers of power electronics applications that are able to operate in challenging environments. 
Understanding the mechanism of radiation tolerance is crucial for further material modification and tailoring of the desired properties. 
In this study, we employ \FD{machine-learning-enhanced atomistic simulations} to assess the stability of both the gallium (\ce{Ga}) and oxygen (\ce{O}) sublattices under various levels of damage.
Our study uncovers the remarkable resilience and stability of the \ce{O}-sublattice, attributing this property 
to the strong tendency of recovery of the $\ce{O}$ defects, especially within the stronger 
disordered regions. 
Interestingly, we observe the opposite behavior of the $\ce{Ga}$ defects that display enhanced stability in the same regions of increased disorder. 
\FD{Moreover, we observe that highly defective 
\bGaO~
is able to transform into \gGaO~
upon annealing due to preserved lattice organization of the $\ce{O}$-sublattice}. 
This result clearly manifests that 
the ultrahigh stability of the $\ce{O}$-sublattice provides the backbone 
for the exceptional radiation tolerance of the $\gamma$/$\beta$ double-polymorphic structure. 
These computational insights closely align with experimental observations, opening avenues for further exploration of polymorphism in $\ce{Ga2O3}$ and potentially in analogous polymorphic families spanning a broad range of diverse materials of complex polymorphic nature.
\end{abstract}
\maketitle

\newpage

\section{Introduction} 

Current Si-based semiconductor technology is facing the fundamental limits of narrow bandgap  
and low breakdown field~\cite{chen2017gan}. 
With the key merits of high breakdown electric field strength, ultrawide bandgap, and high ultraviolet optical transparency, gallium oxide ($\ce{Ga2O3}$) has attracted significant attention as a promising candidate for next-generation semiconductors~\cite{mastro2017perspective, pearton2018review, pearton2018perspective, galazka2018beta, zhang2022ultra, tadjer2022toward}.
Furthermore, the exceptional radiation tolerance of the $\ce{Ga2O3}$ offers promising applications in demanding environments, including space exploration, nuclear power generation, and medical imaging~\cite{bauman2021improving,tak2019gamma,hou2020review,teng2014self}. 
Investigating radiation damage in $\ce{Ga2O3}$ is pivotal for achieving a comprehensive understanding of defect theory and its correlated physical properties, however, it remains a significant challenge.

Recently, experimental studies using ion implantation have revealed that $\beta$-$\ce{Ga2O3}$ maintains its crystallinity up to extremely high fluences of ion irradiation~\cite{azarov2023,azarov2022disorder}, which are much higher than the amorphization thresholds observed in other semiconductor materials, \textit{e.g.}, in $\ce{Si}$~\cite{cheang2001rbs}, $\ce{SiC}$~\cite{debelle2014interplay}, and $\ce{GaN}$~\cite{sequeira2021unravelling}. 
Anber \textit{et al.} and Azarov \textit{et al.} illustrated that the amorphization of $\beta$-$\ce{Ga2O3}$ is notably suppressed by the formation of \gGaO~ phase~\cite{anber2020structural,azarov2022disorder}. 
Huang \textit{et al.}~\cite{huang2023atomic,huang2023atomic2} provided quantitative analysis of scanning transmission electron microscopy (STEM) images, displaying the gradual $\beta$ to $\gamma$ phase transformation progress. 
In most cases, the polymorphic transitions in $\ce{Ga2O3}$ crystals arise from lattice adaptation to the minimum-free-energy state at the given temperature and pressure~\cite{song2020reaction,cora2020situ,machon2006high,castro2020atomic}. 
Ion-beam-assisted atom displacement is a novel approach to reaching metastable configurations, yet it has received limited attention in theoretical investigations. 
In order to understand the radiation effects of this emerging semiconductor material at
the atomic scale, He \textit{et al.}~\cite{he2024threshold} and Tuttle \textit{et al.}~\cite{tuttle2023atomic} have conducted calculations to determine the threshold displacement energy, which is crucial for estimating the extent of radiation damage~\cite{kinchin_displacement_1955,norgett_proposed_1975}.
However, the theoretical study of the ultrafast defect dynamics and crystalline phase transformation instead of amorphization during ion implantation is still insufficiently studied.

Here, we utilize machine-learning-enhanced atomistic simulations and reveal that the crystallinity of the \bGaO~ originates from the resilient lattice structure of the $\ce{O}$-sublattice. 
By integrating classical molecular dynamics (MD) simulations with machine-learning Gaussian approximation potentials (ML-tabGAP)~\cite{zhao2023complex}, we systematically explore how accumulated defects in distinct Ga- and O-sublattices influence the structure of \bGaO~ and examine the recombination behavior of single interstitial-vacancy (Frenkel pair) at various separation distances. 
Furthermore, we apply collision cascade simulations to explore the dynamic evolution of defects under irradiation. 
These findings collectively illustrate the ultrahigh stability of the $\ce{O}$-sublattice and the changeable $\ce{Ga}$-sublattice.
This not only precisely corroborates experimental observations but also uncovers the underlying mechanism behind the remarkable radiation tolerance exhibited by $\ce{Ga2O3}$ materials. 
Moreover, this foundational research paves the way for exploring various polymorphism families and deepens our understanding of defective oxide materials. %

\section{Methodology} 

In this study, we elucidate the recovery mechanisms of \ce{O}-sublattice by performing several different types of atomistic simulations. 
Firstly, we use classical MD simulations to simulate damage buildup in $\beta$-$\ce{Ga2O3}$ via Frenkel pair accumulation (FPA) runs. 
We analyze the FP recombination paths by running simulations with isolated FP defects of different types. 
We also run single-cascade MD simulations to analyze the dynamic effects on damage buildup. 
All MD simulations are conducted using the LAMMPS package~\cite{thompson_lammps_2022} with the recently developed machine-learning (ML) interatomic potential(tabGAP) for $\ce{Ga2O3}$ systems~\cite{zhao2023complex}. 
We also perform defective structure relaxation calculations using DFT structure optimization enabled by the VASP package~\cite{kresse1993ab} to verify the results obtained with the ML-MD method. For visualization of the atomic structures, OVITO~\cite{ovitostukowski2009visualization} is utilized. 
In the following, we describe the technical details of each applied type of simulation separately.

\subsection{Frenkel pair accumulation simulations} \label{sec:fpa_method}

In the FPA simulations, we model the ion-induced accumulation of damage by subsequently inserting randomly distributed FPs into initially perfect \ce{Ga2O3} cells. 
To emphasize the atom-type-specific effects on damage accumulation in $\beta$-$\ce{Ga2O3}$, we generate three 1280-atoms simulation cells, where we insert the FPs of three different types: only Ga, only O, and mixed O/Ga FPs. 
Each FP was subsequently created by displacing a randomly selected atom of the given type (in the O/Ga FPA simulations the $\ce{Ga2O3}$ stoichiometry was maintained) in a randomly selected direction by a vector with the norm randomly selected within the range of 5.8 -- 6.2 \r A. 
The insertion of each FP was followed by a relaxation run to reach the local energy minimum, after that, the system was simulated for 5 ps at 300 K and 0 bar in the isothermal–isobaric ($NpT$) ensemble. 
We also performed longer annealing simulations in the $NpT$ ensemble maintaining zero pressure but raising the temperature from 300 to 1500~K with the rate of 6~K/ps after a given number of FPs was accumulated in the cell. 
Each annealing run was performed for 1 ns, after which the temperature was gradually decreased to 300 K with the same rate of 6 K/ps. 
  
\subsection{\textit{Ab initio} simulations} \label{sec:md-dft_method}
To verify the evolution of damage buildup, which we obtained in our ML-MD FPA simulations, we conducted similar simulations using density functional theory (DFT) methods. 
These calculations were performed using the Vienna \textit{Ab initio} Simulation Package (VASP)~\cite{kresse1993ab}, employing the projected augmented-wave (PAW) method~\cite{blochl1994projector} with 13 $(3d^{10}4s^{2}4p^{1})$ and 6 $(2s^{2}2p^{4})$ valence electrons for Ga and O, respectively. The Perdew-Burke-Ernzerhof (PBE) version~\cite{ernzerhof1999assessment} of the generalized gradient approximation (GGA) was used as the exchange-correlation functional. In this calculation, FPs were generated cumulatively in a 320-atom $\beta$-$\ce{Ga2O3}$ cell as the FPA simulation. 
During the accumulation progress, the cell was relaxed after every 5 iterative implantations of FPs. 
\HR{Meanwhile, the cells undergo structure relaxation with ML potential to attain the local energy minimum, serving as a reference for comparison with the GGA-DFT relaxation.} 
In the GGA-DFT calculations, the electronic states were expanded in plane-wave basis sets with the energy cutoff of 700 eV. Given the large supercell, the Brillouin zone was sampled with only the $\Gamma$ point. 
Gaussian smearing with a $\sigma$-width of $0.03$ eV was used to describe the partial occupancies of the electronic states.
$10^{-6}$ eV and $10^{-2}$ eV/\r A were chosen as the energy and force convergence criteria for the optimization of the electronic and ionic structures, respectively. 
These simulations were conducted separately for three different types of FPs inserting 100 FPs in total in each cell.

\subsection{Simulations of Frenkel pair recombination} \label{sec:FP_recover}

In FP recombination simulations of isolated defects, 
a stable interstitial atom was initially inserted in a 1280-atom $\beta$-$\ce{Ga2O3}$ cell. 
Subsequently, a vacancy was created by removing an atom of the corresponding type from a neighboring site around the interstitial within the specific coordination shell to control the recombination radius of the created FP defect. 
The system was thermally equilibrated for 5 ps under the $NpT$ conditions at 300 K and 0 bar. 
The statistical averages were obtained over 50 independent runs for Ga and O FPs, separately.

\subsection{Simulations for single cascades} \label{sec:single_cascade}

Single cascade simulations were carried out by giving the recoil energy to the atom randomly selected in the middle of the simulation cell as a primary knock-on atom (PKA). 
The direction of the initial momentum was also randomly selected in the three-dimensional space. 
The total number of atoms in each simulation cell was selected to match the recoil energy, so that the initiated cascade did not interact with the thermally controlled border regions. 
Table~\ref{tab:singlecascade} provides the recoil energy values and corresponding numbers of atoms in the simulation cells.

\begin{table}[htbp]
\centering
\caption{Simulation parameters. The variable $E_\mathrm{PKA}$ represents the initial kinetic energy given to PKA, $n_\mathrm{atoms}$ is the number of atoms in the simulation cell, $n_\mathrm{PKA:O}$ is the number of simulations in which PKA is O, $n_\mathrm{PKA:Ga}$ is the number of simulations in which PKA is Ga. }

\label{tab:singlecascade}
\begin{tabular}{m{2cm} m{2cm} m{1.6cm} m{1.6cm}}
\hline
\hline
$E_\mathrm{PKA}$ (eV) & $n_\mathrm{atoms}$ & $n_\mathrm{PKA:O}$  & $n_\mathrm{PKA:Ga}$  \\ 
\hline 
500  & 81,920 & 69 & 61  \\ 
750  & 81,920 & 65 & 64 \\
1,000 & 81,920 & 61 & 63 \\
1,500 & 160,000 & 62 & 49\\
2,000 & 276,480 & 54 & 61 \\
\hline
\hline
\end{tabular}
\end{table}

Periodic boundary conditions were applied in all directions, with the Nos{\'e}-Hoover thermostat~\cite{hoover1985} controlling the temperature along the borders of the simulation cell to mimic the heat dissipation of bulk materials. 
To ensure simulation efficiency and system stability, an adaptive time step was employed. 
Electronic stopping power as a friction term was applied to the atoms with kinetic energies above 10 eV~\cite{nordlund1998defect}. 
The simulation time of the single cascades was 50 ps. 
To guarantee sufficient statistics, parallel simulations were conducted for each recoil energy with varying PKA. The detailed simulation amounts are presented in Table~\ref{tab:singlecascade}.
The point defects formed in collision cascades were identified using the Wigner-Seitz analysis method.

\section{Results} 

\subsection{Stability of cation and anion sublattices in \bGaO} \label{sec:stability}

\begin{figure}[h]
 \includegraphics[width=8.6 cm]{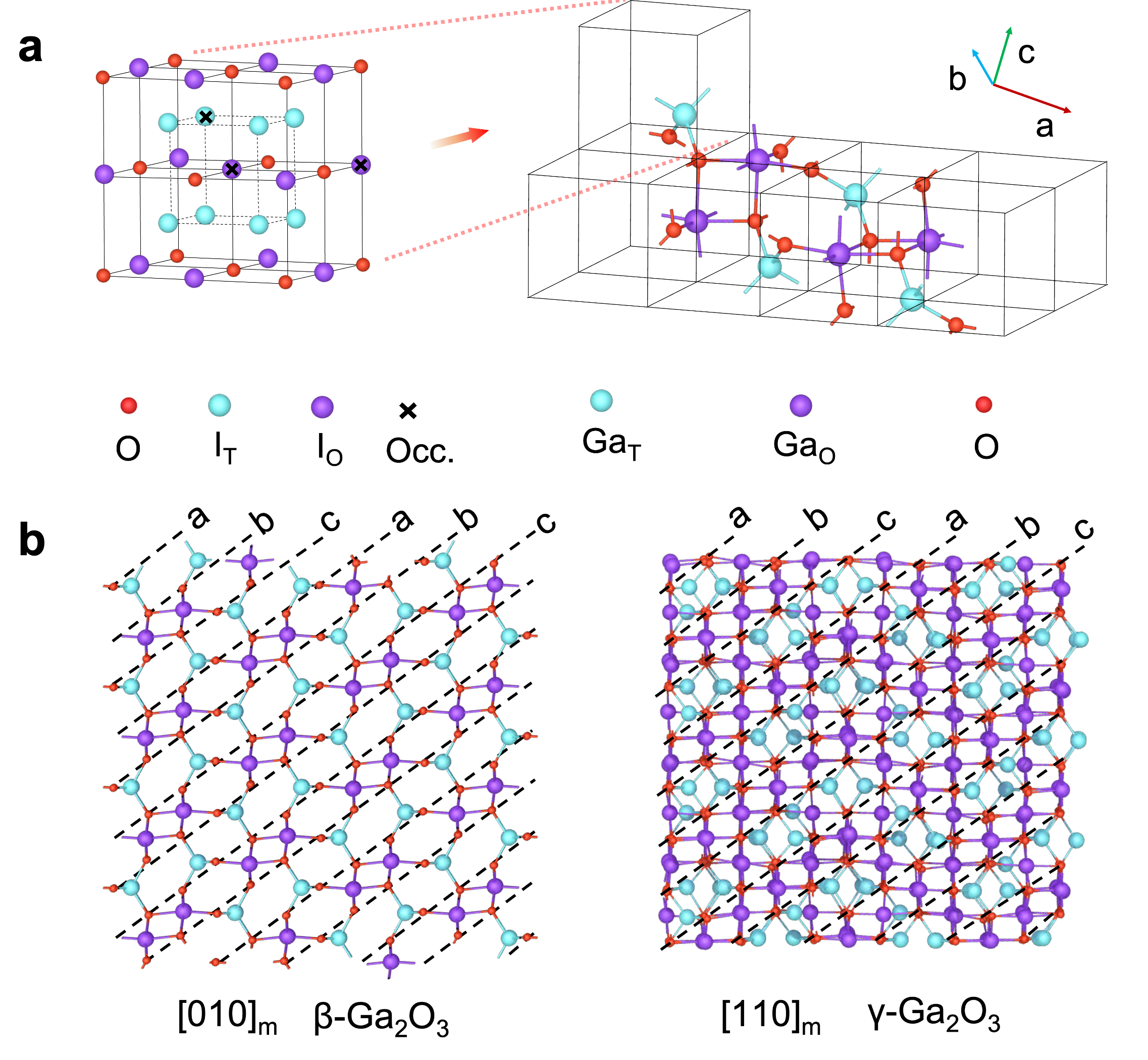}
\caption{(a) Schematic representation of O FCC stacking in the $\beta$-$\ce{Ga2O3}$ unit cell; the right part is the FCC unit cell with tetrahedral (I$_\mathrm{T}$) and octahedral (I$_\mathrm{O}$) interstitials, and the left part is the $\beta$-$\ce{Ga2O3}$ unit cell with O FCC cubic frame. 
Tetrahedral interstitials and Ga$_\mathrm{T}$ are colored purple; octahedral interstitials and Ga$_\mathrm{O}$ are blue; O ions are red; the black cross symbol marks the occupied Ga sites. (b) The snapshots of the pristine $\beta$- (left) and $\gamma$-$\ce{Ga2O3}$ (right) lattice. The dashed line indicate the perfect abc-abc FCC stacking of the O-sublattices.}
 \label{fig:scheme}
\end{figure}

Figure~\ref{fig:scheme} demonstrates the face-centered-cubic (FCC) unit cell of the \ce{O}-sublattice of \bGaO, which is common for both $\beta$- and $\gamma$-phases of \GaO.  
The difference between these two phases is in the location of Ga cations that occupy different tetrahedral (Ga$_T$) and octahedral (Ga$_O$) sites~\cite{geller1960crystal,aahman1996reinvestigation}. 
It has been previously observed that \bGaO~ transforms into the $\gamma$-phase under ion implantation~\cite{garcia2022formation,azarov2023,huang2023atomic,huang2023atomic2}, indicating that the integrity of the $\ce{O}$-sublattice is maintained, while Ga atoms intricately rearrange with the accumulated damage dose. 
In the monoclinic $\beta$-phase, the molar ratio of tetrahedral (Ga$_\mathrm{T}$), octahedral (Ga$_\mathrm{O}$) cations and O anions is $1:1:3$, while within the FCC unit cell, see Figure~\ref{fig:scheme}(a), the ratio of the tetrahedral (I$_\mathrm{T}$), octahedral (I$_\mathrm{O}$) sites, and O lattice sites is $8:4:4$. In Figure~\ref{fig:scheme}(a), we indicate all tetrahedral and octahedral sites potentially available for Ga cations within the O FCC lattice unit cell: the purple balls show the potential Ga$_\mathrm{O}$ sites and the cyan balls the potential Ga$_\mathrm{T}$ sites. 
The \ce{O}-sublattice is shown by the small red balls. 
For clarity, we cross the sites that are occupied by Ga cations in the $\beta$-phase (compare with the conventional image of the \bGaO~ monoclinic unit cell shown in Figure \ref{fig:scheme}(a) to the right). 
As one can see, the majority of the tetrahedral and octahedral sites are unoccupied.
On the other hand, the cubic $\gamma$-phase exhibits the defected spinel structure with partially occupied Ga$_\mathrm{T}$ and Ga$_\mathrm{O}$ in random order, which is challenging to show in a unit cell~\cite{ratcliff2022tackling,playford2014characterization}.

\begin{figure*}[htbp!]
 \includegraphics[width=17.2 cm]{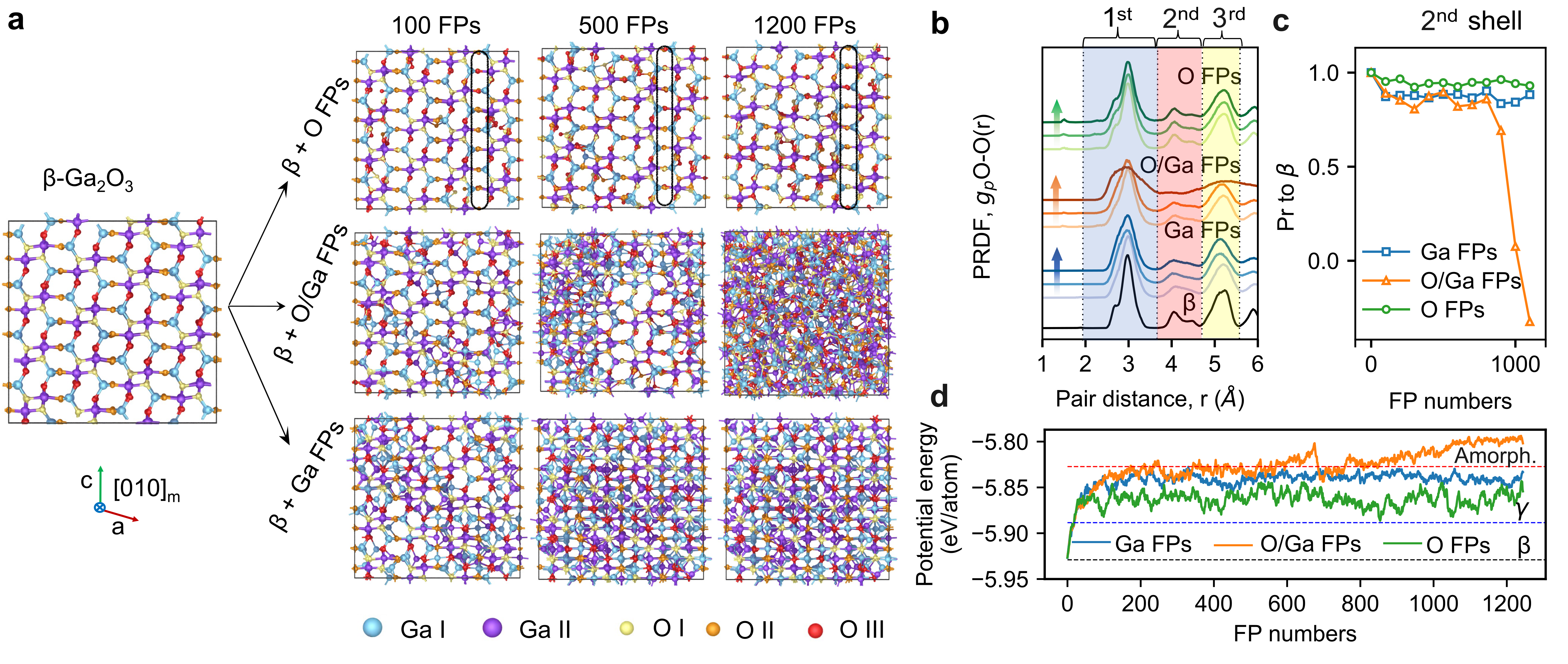}
\caption{Accumulated different types of FPs in $\beta$-$\ce{Ga2O3}$. 
(a) The snapshots of the pristine $\beta$-$\ce{Ga2O3}$ lattices after introducing different types of FP. The five distinct crystallographic sites Ga-I, Ga-II, O-I, O-II, and O-III in the pristine lattices are colored separately in cyan, purple, yellow, orange, and red, respectively.
(b) Analysis of the PRDFs of $\ce{O}$-sublattices with different amounts and types of additional FPs in $\beta$-$\ce{Ga2O3}$ lattices (blue is Ga FPs, orange is O/Ga FPs, green is O FPs, the color intensity increases systematically with the quantity of Frenkel pairs, specifically designated as 100, 500, and 1200 FPs, respectively.).
(c) The Pearson correlation coefficient, Pr, calculated within the 2$^\mathrm{nd}$ shell for the O-O PRDFs of the increasing damaged $\ce{O}$-sublattices concerning the pristine $\beta$-$\ce{O}$ PRDFs as a function of the FP number.
(d) Potential energy of the cell as a function of FP number. The dashed lines correspond to the potential energy of $\beta$, $\gamma$, and amorphous $\ce{Ga2O3}$ at \HR{300 K}, respectively. }
 \label{fig:fpa}
\end{figure*} 

To enable the visual comparison between the two lattices of pristine $\beta$- and $\gamma$-phases, in Figure~\ref{fig:scheme}(b), we analyze the composition of the Ga layers between the abc-abc stacked close-packed planes in the O FCC-sublattice, which are shown by the dashed lines.
In the defective spinel structure of the $\gamma$-phase, one Ga layer is primarily composed of Ga$_\mathrm{O}$, while the other layer contains both Ga$_\mathrm{T}$ and Ga$_\mathrm{O}$, contrasting with the perfectly ordered $\beta$-phase, where Ga$_\mathrm{T}$ and Ga$_\mathrm{O}$ appear alternatively in the different Ga layers.

In our study, we aim to investigate in detail whether the phase transformation in \bGaO~ during ion irradiation depends on the type of the defects (cation or anion).  
We conducted the type-specific FPA simulations to systematically analyze the damage buildup in both sublattices separately as well as the damage buildup in the entire structure without separation into sublattices. Figure~\ref{fig:fpa} illustrates the evolution of the crystal structure of \bGaO~ with the accumulation of the various types of FPs. 
Five different colors are used to show crystallographically different types of two Ga cations (Ga-I, Ga-II) and the three O anions (O-I, O-II, O-III) sites in the pristine lattice to facilitate the analysis of 
atom displacements. 

In the O FPA simulations [top row in Figure~\ref{fig:fpa}(a)], we observe that the $\beta$-$\ce{Ga2O3}$ 
persistently retains its initial phase structure with only a few remaining defects, although the location of some O atoms have changed, compare the colors of O atoms in the outlined regions in the snapshots showing the initial and damaged structures in Figure~\ref{fig:fpa}(a). 
The outlined vertical region shows the same column of the O anions in all three snapshots. 
Initially, in the pristine cell, all sites in this column are consistently colored orange, showing O-II anions. 
With the increase of the accumulation of O FPs, some of these sites are replaced by yellow (O-I) and red (O-III) anions. 
In the Ga FPA simulations [bottom row in Figure~\ref{fig:fpa}(a)], the crystal structure remains highly ordered, however, with the increase of the number of FPs, it gradually resembles more the $\gamma$-phase with denser occupied Ga sites in $[010]$ direction covering the open channels of the $\beta$-phase lattice in the same direction.  

The O/Ga FPA simulations [the middle row of Figure~\ref{fig:fpa}(a)] result in a complete disorder of the $\beta$-phase, although the simulations up to the low damage dose (about 100 FPs) showed similar structural evolution to that in the Ga FPA simulations. 
However, further accumulation of O/Ga FPs (see 500 and 1200 FPs) leads to a rapid crystallinity deterioration, transforming the structure into an amorphous phase. Surprisingly, we see that within the ordered but defective regions that exist in the cells with Ga FPs and O/Ga FPs, the FCC stacking of the O-sublattice is mostly maintained. 
Hence, we plot in Figure~\ref{fig:fpa}(b), the O-O partial radial distribution functions (PRDFs) for the $\ce{O}$-sublattices in the cells, highlighting three coordination shells: the first (2.0–3.6 \r A), the second (3.6–4.6 \r A), and the third shell (4.6–5.6 \r A). 
In the systems where the damage accumulation proceeded separately either within the anion or the cation sublattices, 
we observe that the peaks in the O-O PRDF persist, indicating the anion sublattice structure resists the damage, while in the cells with the accumulation of O/Ga FPs, the main characteristic peaks of the FCC structure disappeared, indicating amorphization of the lattice.

The similarity of the O-O PRDFs of the cells with the increased damage to the O-O PRDF of the pristine \bGaO~ is analyzed in Figure~\ref{fig:fpa}(c), where we plot the Pearson correlation coefficient (Pr) for the corresponding curves in the 2\textsuperscript{nd} shell. 
The O-O PRDFs exhibit a high degree of positive correlation with the $\beta$-O structure in the case of pure O FPs and Ga FPs, demonstrating a significant degree of structural similarity. 
As the accumulation of the mixed FPs, the Pr value initially fluctuates around a rather high value of 0.8, but then it drops abruptly, signifying the robustness of the $\ce{O}$-sublattices in maintaining their structure up to a specific threshold of damage. 
Whereas, surpassing these limits results in the direct collapse of the structure. 
We also analyzed the Ga-Ga PRDFs (see the Supplementary material (SM) Appendix A, Figure~\ref{fig:gardf}) to demonstrate the effect of the damage buildup in specific sublattices on the structural integrity of the $\ce{Ga}$ sublattice. 
In this analysis, we observe that only the formation of Ga and O/Ga FPs leads to the appearance of the $\gamma$-$\ce{Ga}$-like phase while displacing only O atoms does not trigger this transformation.

In Figure~\ref{fig:fpa}(d), we also follow the evolution of the potential energy in all three simulation cells after the insertion of every FP and subsequent relaxation at 300 K. Despite fluctuations of the energy values, the comparison of the three curves clearly illustrates the different effect that the accumulation of different types of FPs causes in the corresponding cells. 
For instance, we see that the potential energy rises the fastest in the cell with the accumulation of the O/Ga FPs, with the highest value exceeding the potential energy of the relaxed amorphous phase.
We also note a sharp drop of this value at about 700 FPs, which exceeds the fluctuation uncertainty bar in the rest of the potential energy curve.
The drop indicates the phase transformation from a highly damaged but still crystalline structure of \bGaO~ to the amorphous state, hence we can conclude that in these simulations we reached the damage tolerance threshold for the \bGaO~ structure before collapsing into the amorphous state. 

\FD{Insertion of Ga FPs brings the potential energy in the simulation cell very close to the amorphous level, however, not exceeding it at any point. We also note that the level of the potential energy in this cell is consistently above the pristine \gGaO~ level, which is the footprint of unstable defects in the gradually damaging lattice.  Since we do not observe any abrupt changes in the value of the potential energy, which would indicate a phase transition, we conclude that the accumulation of the FPs themselves does not yet transform \bGaO~ into a stable \gGaO~ structure and additional annealing of created defects is needed to complete this transformation. Moreover, we notice that the potential energy reaches the saturation level after less than 200 Ga FPs and continues to fluctuate around this value up to the very high doses of 1200 FPs, which we reached in our simulations. This behavior can be interpreted as randomization of the Ga sublattice, however, after it has been randomized the following displacements do not introduce additional disturbance, which would lead to the increase of the potential energy in the cell. }

\FD{The insertion of O FPs introduced the smallest disturbance in the simulation cell, see the lowest green curve in Figure~\ref{fig:fpa}(d). The inevitable accumulation of the defects in this cell brings the potential energy to the higher values than the \bGaO~ and \gGaO~ phases [the corresponding grey and blue dashed lines in Figure~\ref{fig:fpa}(d)]. However, we see much stronger fluctuations in this curve, which indicates stronger relaxation effects in this cell even during the short relaxation runs after insertion of a subsequent O FP. Again, no indication of phase transformation is found in this curve either, while overall the potential energy fluctuates the strongest out of the three curves with barely noticeable increasing tendency. Even at the highest number of FPs (1,200 at the end of the simulation) the potential energy in the cell is not remarkably higher than that after the first 200 O FPs.  The saturation trend observed in both curves for Ga and O FPs separately indicate that the \bGaO~ lattice is capable to accommodate the damage in separate sublattices easier compared to the damage accumulation via displacement of both types of ions at the same time.}

Our results showcase the resilience and stability of the $\ce{O}$-sublattice in \bGaO~ even at very high damage levels. The response of the \bGaO~ to FP type-specific damage accumulation demonstrates that the ordered O-sublattice is the only possible 
structural backbone that is able to preserve the crystallinity of the material exposed to extreme irradiation conditions. 
Our results also indicate that the crystal structure collapses into an amorphous state only after the 
damage in the O-sublattice reaches a critical damage threshold. 
Without the assistance of displaced Ga atoms, the threshold value is very high ($\gg$ 1 dpa) because of the strong recombination trend in the O-sublattice. 
However, with the assistance of Ga FPs, this threshold is lowered more than half ($\sim0.5$ dpa in our simulations). 
\FD{We note here that the damage level in atomistic simulations corresponds to the experimental damage level measured at higher dpa because of the short relaxation times between the FP insertions.}  
\begin{figure*}[htbp!]
 \includegraphics[width=15 cm]{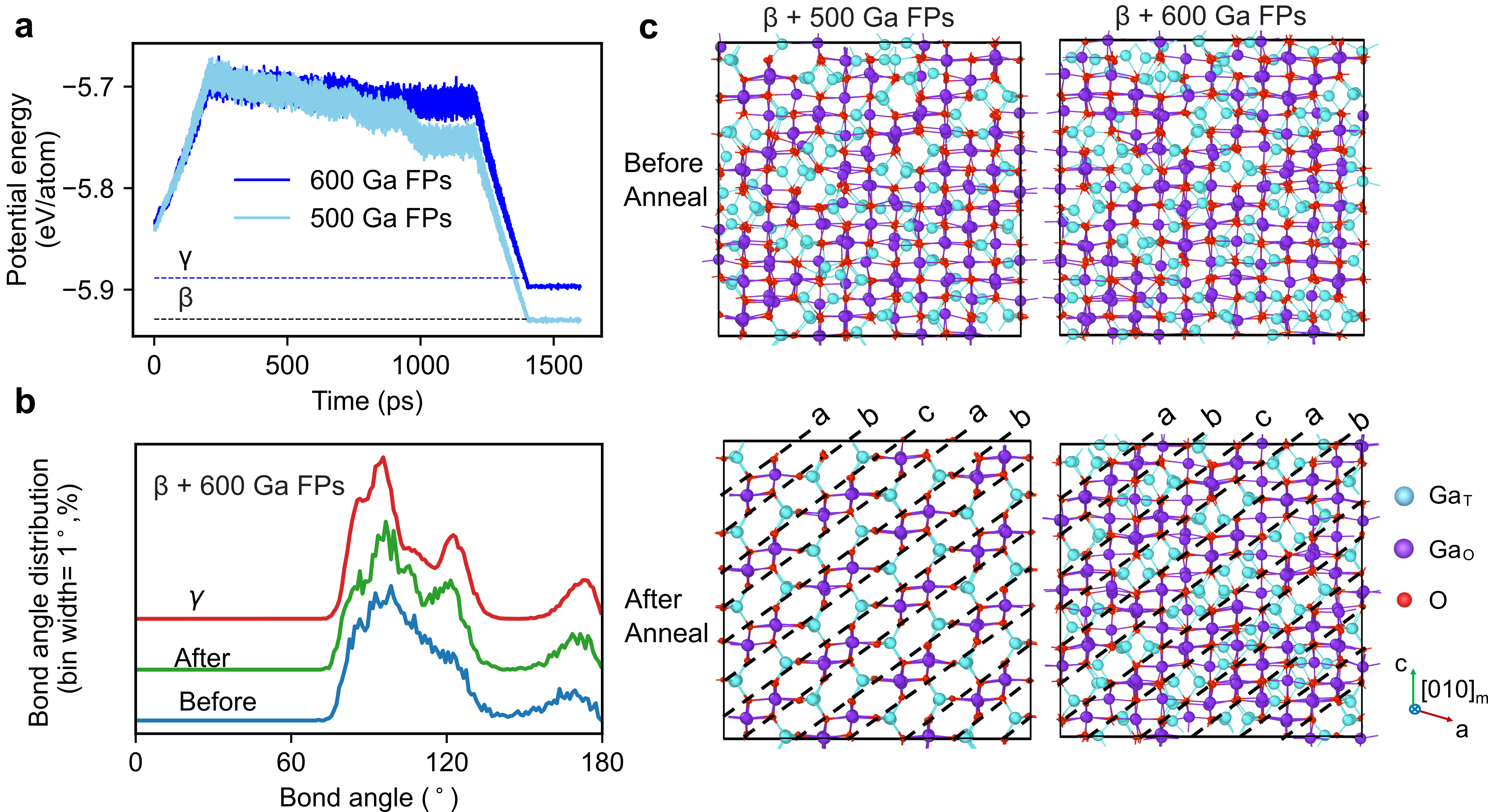}
 \caption{Analysis of the cells with 500 and 600 Ga FPs in 1500 K annealing: (a) Potential energy of the cell during the annealing; the light blue line is for cells with 500 Ga FPs, and the dark blue line is for cells with 600 Ga FPs, the dashed lines correspond to the potential energy of $\beta$- and $\gamma$-$\ce{Ga2O3}$, respectively.
 (b) The bond angle distribution of cells with 600 Ga FPs before (blue line) and after (green line) annealing; the red line is for the pristine $\gamma$-$\ce{Ga2O3}$.
 (c) The snapshots of the cells with 500 and 600 Ga FPs before and after annealing; the dashed line indicates the perfect abc-abc FCC stacking of the $\ce{O}$-sublattices; tetrahedral interstitials and Ga$_T$ are colored purple; octahedral interstitials and Ga$_O$ are blue; O ions are red.} 
 \label{fig: GaFPs_anneal}
\end{figure*}
Moreover in FPA simulations, FPs are introduced randomly within the lattice, which may lead to formation of shallow metastable defects, effectively enhancing the defective state of the lattice not expected as a result of collision cascades. 

To verify the FPA results, we utilized DFT to relax the ML-MD \bGaO~ atomic structures damaged in the same FPA manner, but of a smaller size manageable within the DFT method. 
SM Appendix B, Figure~\ref{fig: DFT_MD}(a) illustrates the very small (below 1 \r A) displacements of the atoms in the lattice after GGA-DFT relaxation of the system obtained in the ML-MD simulations, demonstrating the reliability of the applied ML potential for description of defective structures of $\ce{Ga2O3}$. 
Meanwhile, we compared the potential energy per atom with GGA-DFT relaxation and structure relaxation with ML potential in Figure~\ref{fig: DFT_MD}(b). 
These results show that the ML relaxation agrees closely with the GGA-DFT calculations, further verifying the accuracy and reliability of the ML potential for the FPA simulations.

As mentioned above, the FPA simulations produce highly defective structures, where the transition to \gGaO~ phase is not easy to detect. 
In experiments, the defects generated in cascades may annihilate or relax into energetically more favorable configurations during long time between subsequent cascade events. 
To take into account these relaxation processes, we further applied 
annealing runs at elevated temperature, to promote relaxation dynamics within the MD time span, for 1 ns in the cells with the high damage level (several hundreds Ga FPs). In agreement with our previous results~\cite{zhao2024crystallization}, we also observe that not all defective structures transformed into the \gGaO, but up to a certain 
threshold damage level, the structure collapsed back into the \bGaO. Only 
surpassing the threshold level prompted the lattice to 
undergo a transition into the $\gamma$-phase,
as shown in Figure~\ref{fig: GaFPs_anneal}. We see that the energy curve after the annealing of the 1280-atom-cell with 600 Ga FPs approached the level of the potential energy of the $\gamma$ phase [the dark blue curve in Figure~\ref{fig: GaFPs_anneal}(a)], indicating the phase transformation. 
While the annealing of the cell with 500 Ga FPs resulted in the system energy being very close to that of the $\beta$-phase [the light blue curve in Figure~\ref{fig: GaFPs_anneal} (a)]. 
\FD{Moreover, we draw attention to a step-like feature at $\sim1$ ns in the light-blue curve, which indicates transition from the defective to the perfect $\beta$-phase, which subsequently only cooled down to the level of the perfect \bGaO~ at 300 K.}
The potential energies of the $\gamma$- and $\beta$-phases are shown in blue and black dashed lines, respectively. 
Hence, our theoretical findings reveal that the displaced Ga atoms tend to occupy metastable $\gamma$-Ga sites prompting the lattice to transit into this phase after surpassing the threshold damage level. However, if the large number of Ga atoms still occupy the stable sites of the $\beta$ phase, the lattice does not undergo the transition to a new phase, but inevitably returns back to the original one upon annealing. 

In Figure~\ref{fig: GaFPs_anneal}(b), we show the bond angle distribution for the cell with 600 Ga FPs before and after annealing. For comparison, the same distribution for the pristine \gGaO~ is shown in red.
We see that after the annealing (green curve) the distribution becomes much closer to that of the \gGaO~ structure with the stronger pronounced peak around $120\degree$. 
  
The snapshots in Figure~\ref{fig: GaFPs_anneal}(c) illustrate the lattice structures of the cells with 500 and 600 Ga FPs before (top row) and after (bottom row) the annealing. 
Here we also see that a clear spinel structure~\cite{prins2019location,quintelier2021determination} appears after the annealing of the \bGaO~ with higher number of Ga FPs; we see the alternating cation layers of only \IV{Ga}{O} sites and of a mixture of \IV{Ga}{O} and \IV{Ga}{T} sites. 
It is evident that in the \bGaO~ with fewer Ga FPs (500) the occupation of metastable $\gamma$-Ga sites was insufficient and hence, the structure collapsed back into the initial \bGaO~ lattice. 
The finding confirms that the $\beta$- to \gGaO~ phase transformation requires sufficiently high density of displaced Ga atoms, which occasionally occupy metastable $\gamma$-Ga sites. 
When the occupancy of these sites reaches a threshold value the full recovery back to the \bGaO~ phase requires too many atomic transitions, which increases dramatically the kinetic path. 
Instead, the damaged structure transforms into a metastable, but still energetically favorable \gGaO~ structure during the simulation annealing run (experimentally, the process corresponding to post-irradiation relaxation of the lattice). 
Although some relaxation of radiation-induced defects takes place between the cascades during high-fluence ion irradiation, without thermal assistance the residual defects survive until the next cascade, slowly accumulating until the threshold value for a phase transformation is reached.

\subsection{Recombination of Frenkel pairs}

\begin{figure}[htbp!]
 \includegraphics[width=8.6 cm]{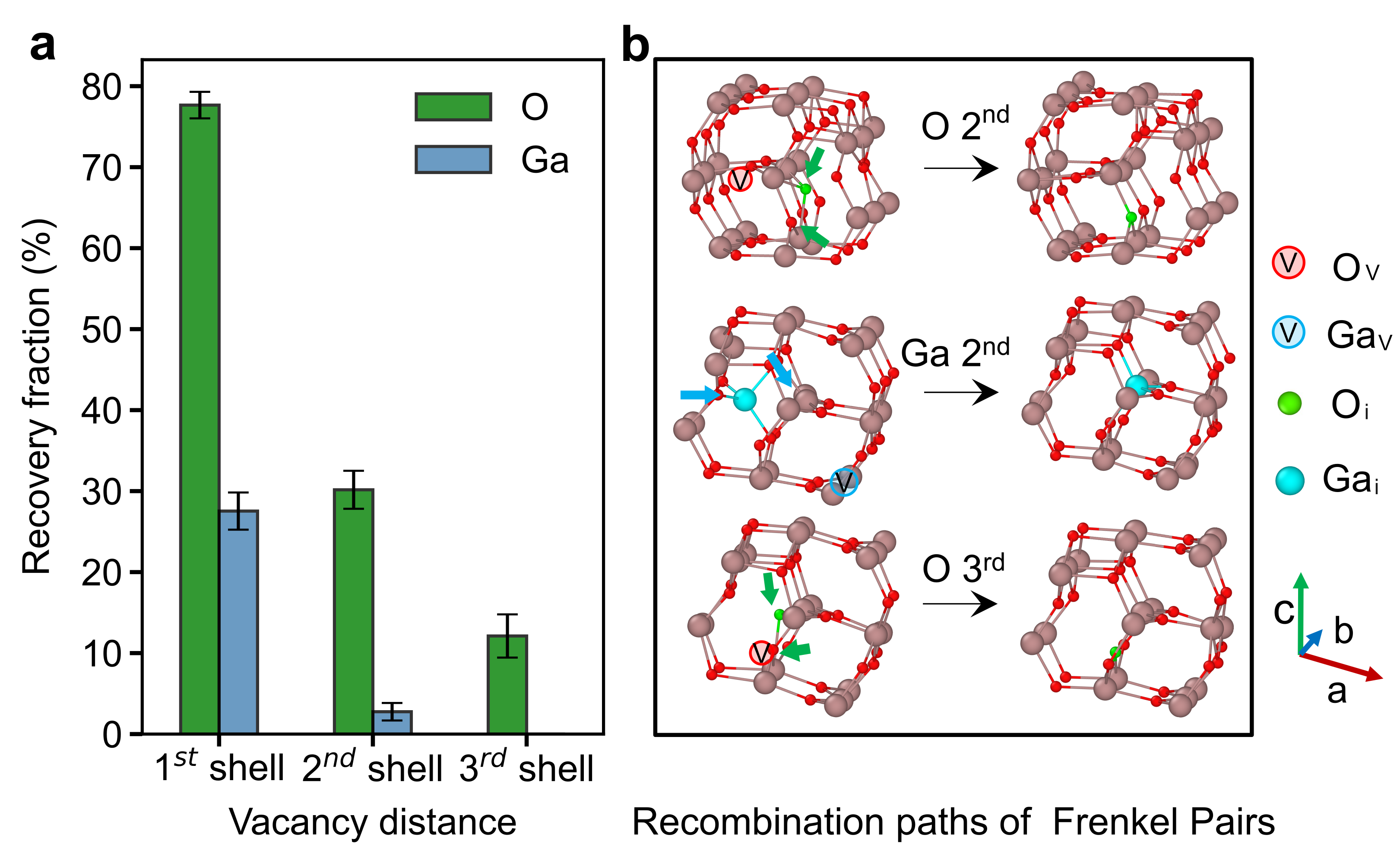}
\caption{Frenkel pair recombination.
(a) Recovery fraction of interstitials (Ga in blue, O in green) to vacancy sites within different coordination shells.
(b) Three representative recombination paths of Frenkel pairs where interstitials recombine with vacancy sites in the 2\textsuperscript{nd} (interstitial O and Ga ions) and 3\textsuperscript{rd} coordination shells (interstitial O ions). } 
 \label{fig:FP_recovert}
\end{figure}

As described in the previous section, we observed in the FPA simulations that it is nearly impossible to accumulate only O FPs in the \bGaO~ lattice. 
This remarkable feature is counter-intuitive bearing in mind the difference in atomic masses of the cation and anion atoms in \GaO.
To investigate the mechanism of such ultrahigh stability of O-sublattice, we inserted a single \IV{O}{i} or \IV{Ga}{i} interstitial into the simulation cell and tested the recovery of the created defect back to a perfect lattice site when a vacancy of the corresponding type appeared in vicinity of the created interstitial. 
We tested different distances from the interstitial to a vacancy placed in random locations around the former but within specific coordination shells. 

Figure~\ref{fig:FP_recovert}(a) shows statistical averages of the recovery percentage of both interstitials with a vacancy of the corresponding type placed in the first, second and third coordination shells away from the interstitial. 
The averaging is done over the relaxation runs of the scenarios where the vacancy changes its position, but within the same coordination shell. 
The error bars show the standard error of the mean. 
The comparison of the recovery fractions for both \IV{Ga}{i} and \IV{O}{i} demonstrates a strong recombination tendency of \IV{O}{i} with vacancies in the first and second coordination shells with the recovery fractions of approximately 80\% and 30\%, respectively. 
\IV{Ga}{i}, on the other hand, recombines with vacancies less efficiently. 
The recovery fraction of \IV{Ga}{i} with a vacancy in the first coordination shell is down to $\sim$30\%, while we found only very few recombinations of \IV{Ga}{i} with a vacancy in the second shell. 
Exemplary recombination paths for \IV{O}{i} and \IV{Ga}{i} are shown in Figure~\ref{fig:FP_recovert}(b). 
It is also noteworthy that \IV{O}{i} is an unstable defect and is eager to give a 'push' to the O atoms in its vicinity, prompting them to occupy the vacant site, while the \IV{O}{i} defect itself recombines with the vacancy newly created by the displaced neighboring O atom. 
Similar remote recombinations with vacancies from the second coordination shell are more rare for \IV{Ga}{i} (see the middle row in Figure~\ref{fig:FP_recovert}(b)). The non-zero recovery percentage was also seen for the \IV{O}{i} with the vacancies placed in the third nearest neighbor positions. The exemplary scenario of such recombination is shown in the bottom row of Figure~\ref{fig:FP_recovert}(b). The observed flexibility in defect recombination of \IV{O}{i} within long distances (up to $\sim6$ \r A) explains the low accumulation rate of O FPs providing strong backbone to maintain the crystallinity of the entire structure. 

\subsection{Collision cascades}
\begin{figure*}[htbp!]
 \includegraphics[width=17.2 cm]{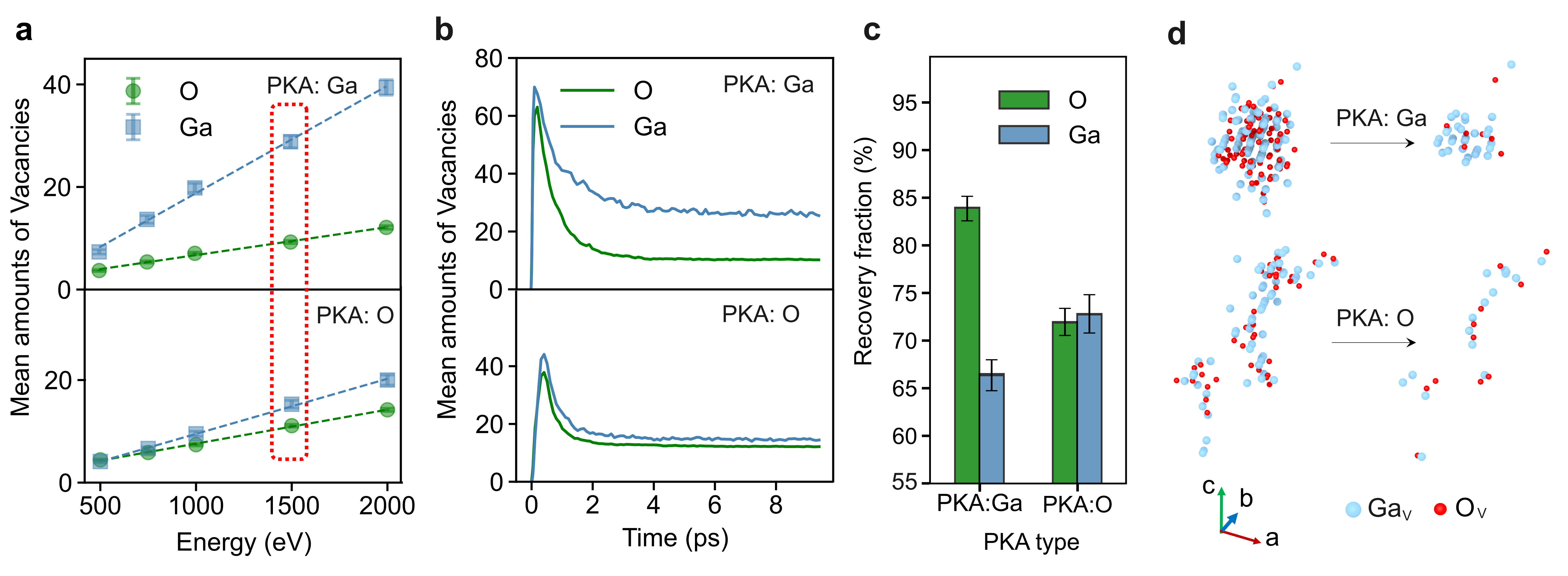}
\caption{Single cascade simulations of $\beta$-$\ce{Ga2O3}$.
(a) Mean number of Ga and O vacancies as a function of PKA energy in single-cascade MD simulations. The upper panel represents Ga PKA, and the lower panel represents O PKA. Vacancies resulting from a recoil energy of 1500 eV are emphasized with the red outline.
(b) The dynamic changes in Ga and O vacancy sites over time in $\beta$-$\ce{Ga2O3}$ during 1500 eV single-cascade MD simulations. The upper panel represents Ga PKA, and the lower panel represents O PKA.
(c) Recovery fraction of O and Ga vacancies in different PKA simulations.
(d) Snapshots of the vacancy sites (red dots are \IV{O}{V} and light blue dots are \IV{Ga}{V}) generated during a 1500 eV single-cascade MD simulation, with defects at both maximum damage and the final primary damage at the end of the simulation. The upper panel represents Ga PKA, and the lower panel represents O PKA. }
 \label{fig:single_cas}
\end{figure*}

Radiation effects in materials are conventionally studied by MD simulations of collision cascades initiated by a randomly selected PKA within the simulation cell~\cite{nordlund2018primary}. 
Unlike the FPA simulations, defects in collision cascades are generated naturally, reducing the probability of the creation of high-energy unstable defects.
In our previous work~\cite{azarov2023}, we have observed the enhanced recrystallization of the $\ce{O}$-sublattice in single cascade simulations in \bGaO. 
In the present study, we analyze the evolution of the number of defects in both Ga- and O-sublattices during collision cascades triggered by different type PKA with several different PKA energies. 
In Figure~\ref{fig:single_cas}(a), we plot the number of vacancies (both types \IV{O}{V} and \IV{Ga}{V}) created by Ga and O PKA separately as a function of PKA energy. 
We see the linear dependence of this relation with the slope depending on the vacancy and PKA types.
While the O PKA produces approximately the same amount \IV{O}{V} and \IV{Ga}{V} almost at all PKA energies (with a slightly higher slope for \IV{Ga}{V}, though), we see much larger difference in the number \IV{O}{V} and \IV{Ga}{V} created by Ga PKAs. 
The growth of \IV{Ga}{V} with PKA energy is much faster than that for the \IV{O}{V} defects. 
Morever, by comparing the slopes of \IV{O}{V} growth created by O and Ga PKA, we observe a surprising trend of even slower growth of \IV{O}{V} in the latter case. In this graph, the error bars show the standard errors of the mean calculated over the simulations, which were performed with the different seed numbers for random selection of the initial position and the direction of the PKA atom in the middle of the simulation cell. 

In Figure~\ref{fig:single_cas}(b), we illustrate the temporal evolution of vacancy defects during the 1500 eV cascade. 
It is clear that both Ga and O PKAs generate similar number of \IV{O}{V} and \IV{Ga}{V} during the ballistic phase of the cascade, but the recombination of \IV{O}{V} defects proceeds much more efficiently, in particular in the cascades triggered by Ga PKAs.
The O PKA generates somewhat higher number of \IV{Ga}{V} which, however, are formed by the atoms that were not displaced sufficiently far from their original positions, resulting in nearly as efficient recombination of this defects as of \IV{O}{V}. 
This difference is further emphasized in Figure~\ref{fig:single_cas}(c), where the recovery fractions in per cent are shown for Ga and O PKA, respectively.
Here, the bars indicating the recovery fraction for \IV{O}{V} and \IV{Ga}{V} in case of the O PKA are similarly above 70\%, while the recovery fraction for \IV{O}{V} created by the Ga PKA reaches nearly 85\%, while almost the same recovery for \IV{Ga}{V} is only slightly above 65\%. We explain this behavior by the different shapes of cascades triggered by the two types of PKA, see Figure~\ref{fig:single_cas}(d), where only the vacancy sites produced throughout the collision cascades are shown. 
\FD{Light O PKAs transfer energy less efficiently in collisions with the lattice atoms, hence, the cascades triggered by these PKAs are widely spread with lower density of the displaced atoms. At the same time, the energy transferred by a Ga PKA is able to generate occasional localized thermal spikes, which contribute to the enhancement of recombination rates of the point defects. }

\section{Conclusions} \label{sec:Conclusions}

In the present study, we show that the stability of the $\ce{O}$-sublattice in \bGaO~ plays the predominant role in preserving the crystallinity under high fluence ion irradiation.
The ultrahigh stability of the $\ce{O}$-sublattice is attributed to its robust FCC stacking structure, which facilitates strong recombination dynamics between interstitial and vacancy defects with a higher likelihood of recovery in the presence of denser defect concentrations. 
In contrast, Ga ions more flexibly occupy different tetrahedral and octahedral interstitial sites available within the FCC $\ce{O}$-sublattices and hence do not preserve or recombine to their original \bGaO~ structure as strongly. 
With increasing damage, Ga ions exhibit reduced preference for sites within the $\beta$-$\ce{Ga}$ sites, which eventually causes the transition to \gGaO~ phase at high-dose damage accumulation. 
Our results demonstrate the extraordinary stability of $\ce{O}$-sublattices, elucidating the micro-mechanism governing the radiation resistance of $\ce{Ga2O3}$, providing insights for material modification and practical applications in the future.

\subsection*{Acknowledgements}

M-ERA.NET Program is acknowledged for financial support via GOFIB project (administrated by the Research Council of Norway project number 337627 in Norway, the Academy of Finland project number 352518 in Finland, and the tax funds based on the budget passed by the Saxon state parliament in Germany).
J. Zhao acknowledges the National Natural Science Foundation of China under Grant 62304097; Guangdong Basic and Applied Basic Research Foundation under Grant 2023A1515012048; Shenzhen Fundamental Research Program under Grant JCYJ20230807093609019.
Computing resources were provided by the Finnish IT Center for Science (CSC).

\bibliographystyle{apsrev4-2}
\bibliography{main}











\newpage
\section*{Supplementary Material}
\renewcommand{\thefigure}{S \arabic{figure}}
\subsection*{Appendix A: The PRDFs of $\ce{Ga}$-sublattices in FPA simulations}

\setcounter{figure}{0}
\begin{figure}[h]
 \includegraphics[width=8.6 cm]{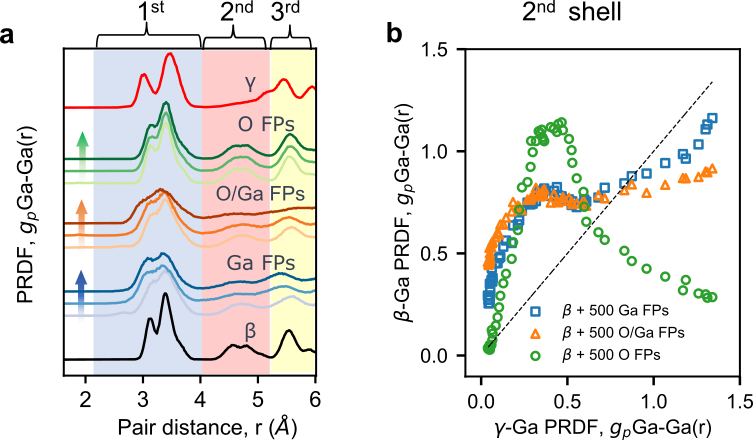}
 \caption{Analysis of the PRDFs of $\ce{Ga}$-sublattices with different amounts and types of additional FPs in $\beta$-$\ce{Ga2O3}$ lattices: (a) Ga-Ga PRDFs for the pristine $\beta$-$\ce{Ga2O3}$ lattice (down); Ga-Ga PRDFs for the pristine $\gamma$-$\ce{Ga2O3}$ lattice (up); blue lines for Ga FPs, orange lines for O/Ga FPs, green lines for O FPs, the color intensity increases systematically with the quantity of FPs, specifically designated as 100, 500, and 1200 FPs, respectively. (b) The similarity of the PRDF values of the $\beta$-$\ce{Ga2O3}$ with 500 different types of FPs versus the PRDF of the pristine $\gamma$-$\ce{Ga}$ within the $2^\mathrm{nd}$ shell.} 
 \label{fig:gardf}
\end{figure}

In Figure~\ref{fig:gardf}(a), we analyze the changes in the Ga-Ga partial radial distribution functions (PRDFs) concerning the accumulation of Frenkel pairs (FPs) of a specific type. Specifically, Ga-Ga PRDF features are examined separately within the $1^\mathrm{st}$ (2.2 -- 4.0 \r A) and $2^\mathrm{nd}$ shell (4.0 -- 5.2 \r A). By comparing the PRDFs of the pristine $\beta$- and $\gamma$-$\ce{Ga2O3}$ lattices, a distinct feature within the $2^\mathrm{nd}$ shell is noticeable only in $\beta$-$\ce{Ga2O3}$, showing peaks at approximately 4.5 \r A. With an increasing number of Ga FPs in a pristine $\beta$-$\ce{Ga}$ sublattice, this feature gradually disappears, and the PRDF of the $\beta$-$\ce{Ga}$ approaches the shape of the $\gamma$-$\ce{Ga}$. When introducing only O FPs, the peaks within the second shell remain visible, even with 1200 O FPs. If FPs are added randomly without considering the ion type, the feature also gradually diminishes. To quantitatively analyze the observed similarities in the curves, Figure~\ref{fig:gardf}(b) plots the PRDF values in the second shell of the damaged $\beta$-$\ce{Ga}$ with 500 FPs of three different types—pure O, Ga, and a mixture of O/Ga FPs—against the PRDF of the pristine $\gamma$-$\ce{Ga}$ within the same interatomic distance range.
In Figure~\ref{fig:gardf}, we observe that the PRDF of the damaged lattice with Ga FPs and mixed FPs is similar to the $\gamma$-$\ce{Ga}$, while the $\ce{Ga}$-sublattice maintains its distinct structure with only O FPs. Combining with the snapshots in Figure~\ref{fig:fpa}(a), the damaged structure with mixed FPs gradually undergoes amorphization, suggesting that high-dose randomly displaced Ga ions transition the $\beta$-$\ce{Ga}$ lattice to defective $\gamma$-$\ce{Ga}$ sites or amorphous Ga sites.\par

\subsection*{Appendix B: ML-MD \textit{vs.} DFT simulations}

\begin{figure}[h]
 \includegraphics[width=8.6 cm]{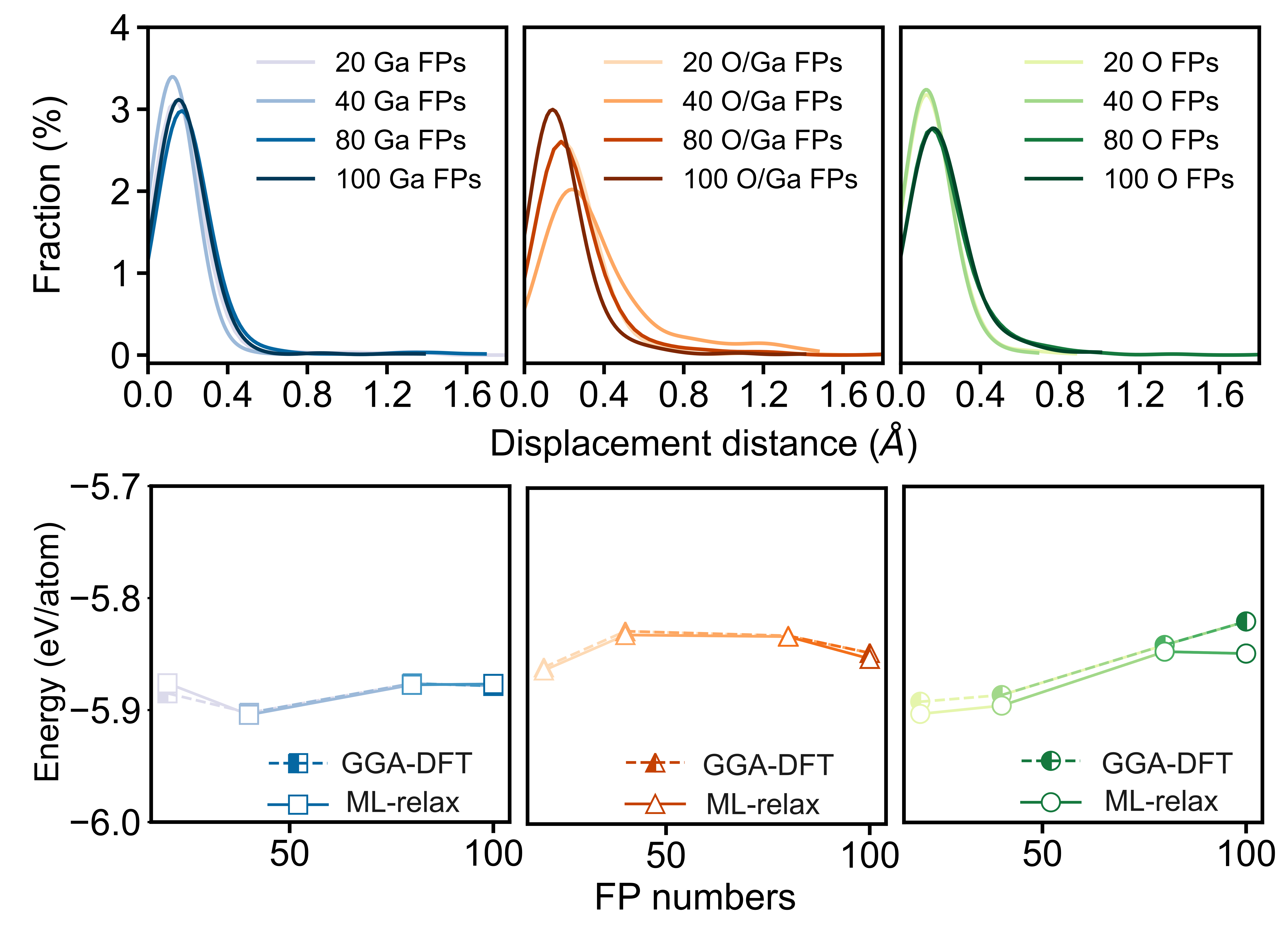}
 \caption{The validation of the FPA simulation through DFT calculation:
 (a) Analysis of the atomic displacement distance after GGA-DFT calculation for the defective \bGaO~ with various O (green), O/Ga (orange), and Ga (blue) FPs. (b) The potential energy of frames with various O (green), O/Ga (orange), and Ga (blue) FPs after GGA-DFT calculations and structure relaxation with ML potential. The intensity of the color increases with the number of FPs. } 
 \label{fig: DFT_MD}
\end{figure}

The precision of molecular dynamics (MD) simulations depends on the choice of the potential employed. In this context, we designed a synergistic approach, combining GGA-DFT calculations with MD simulations, to verify the accuracy of the machine-learning (ML) potential we utilized. As plotted in Figure~\ref{fig: DFT_MD}(a), the ion displacements after GGA-DFT relaxation predominantly remain below 1.0 \r A for various types of FPs. This observation demonstrates the rational and dependable generation of the damaged structure in FPA simulations. Examining the potential energy of frames with different FPs (Figure~\ref{fig: DFT_MD}(b)), we observe that the potential energy obtained from structure relaxation simulations with ML potential closely approximates the GGA-DFT results, with a single exception found in the case of the frame with 100 O FPs. This consistency in energy level demonstrates the ML potential's capability to accurately describe the defective structure in FPA simulations.  The exception point may arise from small differences in the GGA-DFT and ML potential energy surfaces that drive the structure relaxation to different local minima. In general, The DFT calculations exhibit a high level of consistency with our FPA simulations, verifying that the ML potential provides accurate predictions for the damaged structure in $\ce{Ga2O3}$.

\end{document}